\documentclass[twocolumn,showpacs,preprintnumbers,amsmath,amssymb]{revtex4}
\usepackage{graphicx}
\usepackage{dcolumn}
\usepackage{bm}

\begin{document}

\preprint{}

\title{Atomistic modeling of the electrostatic and transport properties of a simplified
nanoscale field effect transistor}

\author{Li-Na Zhao}
\author{Xue-Feng Wang}
\email[corresponding author: ]{xw@atomistix.com}
\author{Zhen-Hua Yao}
\author{Zhu-Feng Hou}
\author{Marcus Yee}
\affiliation{Atomistix Asia Pacific Pte Ltd, Unit 106, 16 Nanyang Drive, NTU
Singapore 637722}
\author{Xing Zhou}
\author{Shi-Huan Lin}
\author{Teck-Seng Lee}
\affiliation{School of Electrical and Electronic Engineering, Nanyang Technological University,
Nanyang Avenue, Singapore 639798}
\date{\today}

\begin{abstract}
A first-principle model is proposed to study the electrostatic
properties of a double-gated silicon slab of nano scale in the
framework of density functional theory. The applied gate voltage is
approximated as a variation of the electrostatic potential on the
boundary of the supercell enclosing the system. With the electron
density estimated by the real space Green's functions, efficient
multigrid and fast Fourier Poisson solvers are employed to calculate
the electrostatic potential from the charge density. In the
representation of localized SIESTA linear combination of atomic
orbitals, the Kohn-Sham equation is established and solved
self-consistently for the wavefunction of the system in the local
density approximation. The transmission for ballistic transport
across the atomic silicon slab at small bias is calculated. The
charge distribution and electrostatic potential profile in the
silicon slab versus the gate voltage are then analyzed with the help
of the equivalent capacitive model. Quantum confinement and short
gate effects are observed and discussed.
\end{abstract}

\maketitle

\section{Introduction}
The size of Si based metal-insulator-semiconductor (MIS) devices has
been shrunk aggressively to nanoscale nowadays \cite{roadmap} in
design of very large scale integrated circuits. The
quantum-mechanical nature of transport becomes inevitably prominent
and there are extensive efforts devoted to modeling quantum
behaviors in nano-MIS devices based on the effective mass
approximation or the empirical tight-binding models.
\cite{S.Datta.1,J.Crofton, H.Wu,A.Rahman,
E.Fuchs,S.Ahmed,L.F.Register,E.Polizzi,A.Pecchia} Besides the
traditional top-down technologies, bottom-up techniques has also
been developed to fabricate Si nanostructures such as nanowires for
future nanodevices. \cite{Y.Cui} As a result, the first-principle
study on electronic and transport properties of atomic Si systems
becomes attractive. \cite{A.Pecchia1,J.M.Soler}

Due to the extensive application of Si based field effect
transistors, some efforts have been carried out to the
first-principle understanding of their behaviors in the past years.
Evans et al. \cite{M.H.Evans} have studied the effect of Si-SiO$_2$
interface roughness on the electron mobility in a Si based MIS
structure by describing the roughness by first principles. Based on
the above method, Hadjisavvas et al. \cite{G.Hadjisavvas} have
proposed an explanation for electron mobility enhancement in
strained MIS field effect transistors (FETs). Fonseca et al.
\cite{Fonseca} and Liu et al. \cite{L.Liu} have established a
two-terminal system of Si-HfO$_2$-Si and Al-SiO$_2$-nSi respectively
described by a self-consistent density functional theory (DFT) and
studied the transport properties through the HfO$_2$ and SiO$_2$
slab in the framework of the nonequilibrium Green's function (NEGF).
In this way, they have estimated the leakage current through the
ultra-thin oxide barriers of MISFETs as a function of the voltage
drop. Landman et al. \cite{U.Landman} have studied the quantum
transport through short Si nanowires passivated by hydrogen atoms
and attached between Al electrodes by solving the eigenchannels of
the scattering states in the framework of the density functional
method with plan wave basis. Similarly, Ng et al. \cite{Ng} studied
the quantum transport through hydrogenated Si nanowires with Li
electrodes employing the NEGF-DFT approach integrated in the
Atomistix ToolKit (ATK) package. Fern\'{a}ndez-Serra et al.
\cite{M.V.Fernandez-Serra} have investigated the effect of doping on
electronic quantum transport in Si nanowires by assuming the same
material (Si) for electrodes and the device regions. Later on,
Markussen et al. \cite{T.Markussen} have continued the work for
longer wires and studied the crossover from ballistic to diffusive
transport based on the DFT and a recursive Green's function
approach. Besides the above two-terminal models, three terminal
models have also been used for Si systems to take into account the
gate effect. Using the NEGF-DFT approach with the help of ATK
package, Dai et al. \cite{Z.X.Dai} have investigated the transport
properties of a Si$_4$ cluster sandwiched between two Al electrodes.
By assuming a constant potential shift in the molecular region
\cite{J.Taylor.1,S.H.Ke} when a gate voltage is applied to the third
terminal, they have observed charge transfer from the molecular and
transconductance oscillation.

The aim of this work is to go further from small three-terminal
molecular systems to more realistic multi-terminal MIS devices of
nanoscale using the NEGF-DFT approach. However, a rigorous treatment
\cite{S.Datta} of realistic multiterminal systems is computationally
too expensive. Fortunately, particle exchange between the gate and
the channel is usually very limited as they are separated by the
insulator slab and the gate terminal works as a controller of
electrostatic potential in the device region. To assure the
simulation feasible with the present computation capability while
catching the most important characteristics of nano MISFETs, we use
a simplified model and focus ourselves on the gate effects on the
electrostatic and transport properties under a small source-drain
bias (in the linear region). As we know, in a real MIS device, how
the gate voltage is distributed into the channel of the device or
the electrostatic property of the device as a MIS capacitor is
critical in determining the characteristics of the whole device.
\cite{N.Arora} For example, double-gate (DG) structure has been
considered as the most promising device geometrical structure for
nano MIS devices due to its good electrostatic integrity
\cite{F.G.Pikus, G.Baccarani}. In our model, we will take the effect
of gate voltage distribution into account  instead of simply adding
a constant shift to the electrostatic potential. The gate voltage
that controls the boundary condition and the corresponding variation
of potential inside the device is self-consistently determined by
Poisson equation.

The rest of the paper is organized as follows. In Sec. II, we define the device model and
briefly describe the NEGF-DFT formalism and the computation method. In Sec. III, the result for
electrostatic properties, charge and potential, are reported and analyzed with the help of
density of states (DOS) and transmission spectra. Finally, a brief summary is given in Sec. IV.

\section{model and method}

\subsection{Device Model}

\begin{figure}
\begin{picture}(300,240)
\put(0,40){\includegraphics{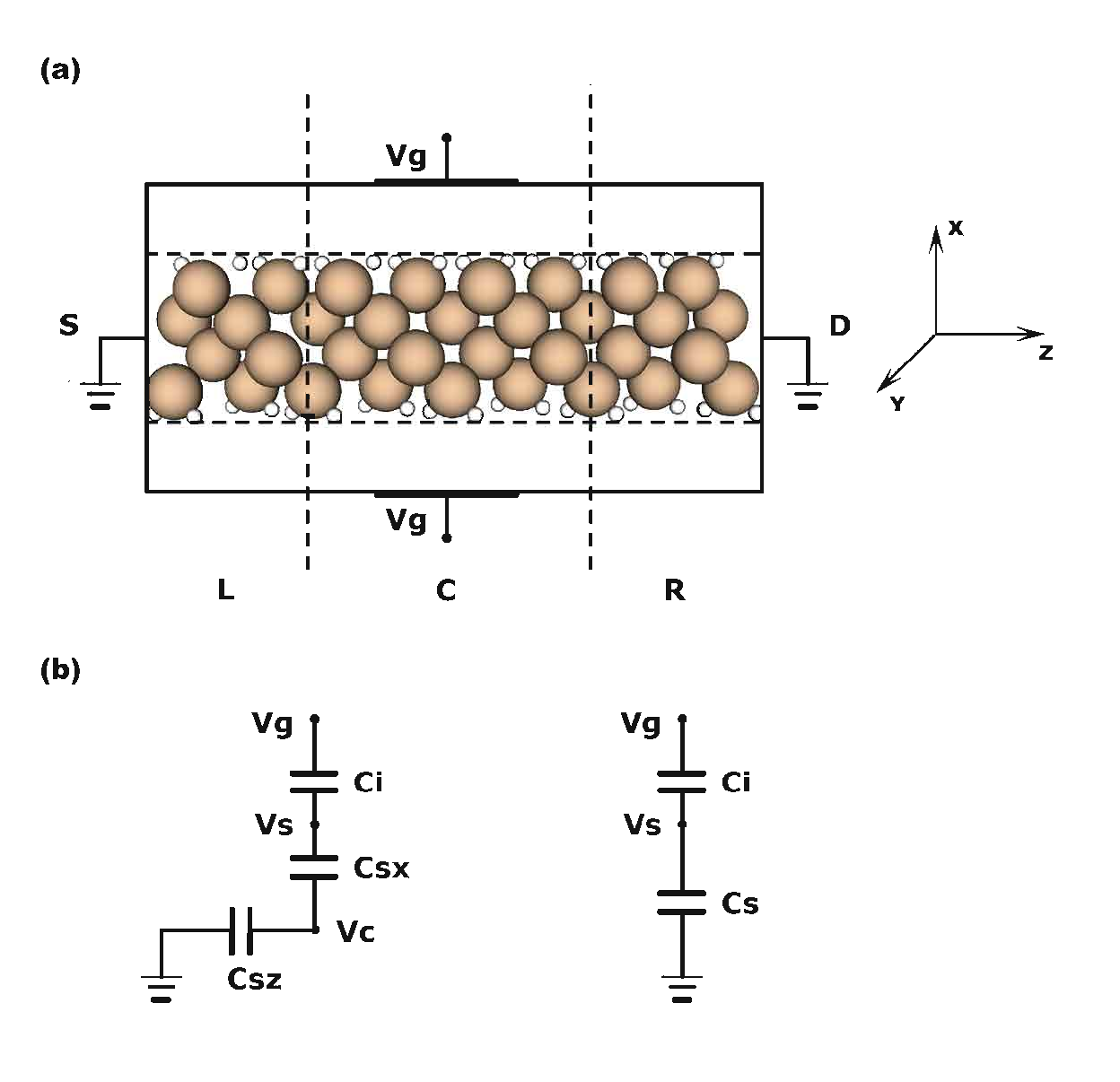}}
\end{picture}
\protect\caption{(a) Atomistic scheme of a model double-gate MISFET
enclosed in one supercell of length $21.72$ \AA\ (four Si unit
cells). The system has the left electrode (L), the central region
(C), and the right electrode (R) separated by the vertical dashed
lines at $z=5.43$ and $16.29$ \AA. The atomic positions projected to
the $x$-$z$ plane are illustrated by the big (Si) and small (H)
spheres. The short thick bars (marked by the gate voltage $V_g$)
indicate the position of the gates. The gate region is the supercell
region between the two bars with the gate length along $z$. (b) The
two equivalent capacitive circuits for the double gate MISFET. The
surface potential $V_s$ is the averaged potential change on the
outer surfaces of the H slabs (indicated by the horizontal dashed
lines in (a)) in the gate region when a gate voltage is applied. The
central potential $V_c$ is the averaged potential change on the
central $y$-$z$ plane in the gate region.} \label{fig1}
\end{figure}

We consider a prototype DG MISFET as schematically shown in
Fig.~\ref{fig1}(a). The device is made from a Si slab with thickness
of nm order in the $x$ direction. The channel ($z$) direction is
chosen along Si crystal $[001]$ orientation and the interfaces along
$(100)$ which is the preferred direction due to lower interface
defect density and higher mobility. H atoms are used to passivate
the dangling Si bonds on the surfaces. The H-Si bond length is
relaxed and has a value of 1.49 \AA\ while the Si-Si bond length is
2.35 \AA. Further on either side of the Si slab a metal gate of
variable length along $z$ direction is located a few angstroms away
from the H atom sheet, as indicated by the thick solid bars in
Fig.~\ref{fig1}(a). Note that here we insert the vacuum slabs,
instead of oxide slabs, between the gates and the Si slab as the
insulator. This is one of the simplifications introduced in this
model in order to reduce the computation load. Since the dielectric
constant of vacuum ($\kappa_v=1$) is much lower than that of oxide
(e.g. $\kappa_{ox}=3.9$ for SiO$_2$), we expect a lower insulator
capacitance (between the gate and the channel) in our model and,
hence, a larger gate threshold than that for devices with oxide
insulator slabs.

Four different model structures of total length 21.72 \AA\ are
studied in this paper and they are denoted as $S_{mn}$ for models
with short gate (length $5.43$ \AA) and $L_{mn}$ for the model with
long gate (in the full central region of length $10.86$ \AA). The
index $m$ ($n$) is used to distinguish models with different
thickness of the Si slab (vacuum slab). The Si slab in the models is
$m$ unit-cells thick ($5.43$ and $10.86$ \AA\ for $m=1$ and $2$
respectively). The corresponding vacuum slab thickness is $1.9$ and
$2.9$ \AA\ for $n=1$ and $2$ respectively.

Another simplification is to approximate the gate effect with a
boundary condition based on the following physical consideration.
When we apply a small voltage drop between the Si slab and the metal
gates we shift the Fermi energy in the latter by letting the former
grounded. Due to the high electron density in metal, the
electrostatic potential in the gates will shift in parallel with the
gate Fermi energy or the voltage drop. As a result, the role of the
metal gates on the channel is to control the electrostatic boundary
of the supercell as enclosed by the solid lines in
Fig.~\ref{fig1}(a). Understanding this gating mechanism, we can
exclude the gate materials from our model system (inside the
supercell) while taking its most important effect into account by
shifting the boundary condition as further explained in next
section.

A third simplification is to use the same material for the source/drain electrodes and the
channel. In real MISFET devices, the drain and source electrodes are doped and the contact
between the electrodes and the channel adds to the complexity of the system and makes the result
difficult to analyze. In this work, we focus on the gating effect in the channel region and it
would be more practical to handle a simplified system. This simplification is expected not to
affect the conclusion when the bias is low and has been used by other authors.
\cite{M.V.Fernandez-Serra}

After the above simplifications, in our model, the double-gated
MISFET is a Si slab with known electrostatic boundary and can be
studied by the NEGF-DFT method. The system is composed of a hydrogen
passivated Si slab and two vacuum slabs. Along $z$ axis, as
separated by the vertical dash lines in Fig.~\ref{fig1}(a), the
system is divided into three regions: the left electrode (L) the
central region, or the channel region (C), and the right electrode
(R). In $z$ direction, the electrostatic potential of a Si slab
without gate is used as the boundary condition of the electrostatic
potential on the left and right boundaries of the supercell. This
implies that the left and right electrodes are assumed
semi-infinite. In $y$ direction, the periodic boundary condition is
applied to assume an extending system in this direction. In $x$
direction, the boundary is variable depending on the system size and
the gate voltage and will be specified in the corresponding context.
Furthermore, we assume a uniform Fermi level inside the supercell.
This approximation is justified in case of small bias between the
source and drain electrodes because the gate voltage induced
variation of Fermi energy happens mainly in the vacuum slab and the
density of states there is negligible.

\subsection{Theoretical Formalism}

To analyze the electrostatic and transport properties of the atomic scale MISFET, we utilize
density functional theory (DFT), \cite{P.Hohenberg, W.Kohn}
which converts the many-body system into a single-particle system, and the two-terminal
nonequilibrium Green's function (NEGF) technique \cite{A.P.Jauho, S.Datta},
which takes care of the coherent transport through the system between the source and the drain
electrodes in Landauer picture \cite{M.Buttiker, S.Datta, J.Taylor,M.Brandbyge,note}.
With standard norm-conserving pseudopotentials \cite{D.R.Hamann} to describe the effective
interaction of the valence and core electrons, we construct
a localized SIESTA linear combination of atomic orbitals (LCAO) basis set $\{ \psi_i \}$
\cite{P.Ordejon} as the representation for expanding the electronic wavefunctions.
According to the Hohenberg-Kohn theorem, \cite{P.Hohenberg} in the Born-Oppenheimer
approximation, the fully interacting electron problem for the ground state can be mapped into a
variational problem in terms of a single-particle density $\rho(\bm{r})$.
The Kohn-Sham (KS) equation \cite{W.Kohn} is then established for the atomic system
in the representation $\{ \psi_i \}$ with the corresponding KS Hamiltonian operator expressed in
matrix form as $H[\rho(\bm{r})]$.
In extending systems, we use the supercell technique with the periodic boundary condition to
define and solve the KS equation in the whole real space.
The dimension of the supercell is chosen larger than the screening length of the KS potential
in any direction so the real system is well mimicked by that inside the supercell.
In open systems, the wavefunction differs from that in closed system and this is taken
care of by using the Green's function of a semi-infinite lead \cite{S.Sanvito} as the boundary
condition
for the Green's function inside the supercell.
With the KS Hamiltonian
$H$ we calculate the retarded (advanced) Green's function (GF)
$G^{R(A)} (\varepsilon)$, Keldysh GF $G^< (\varepsilon)$ and retarded (advanced)
self-energy $\Sigma^{R(A)} (\varepsilon)$. The density matrix $\rho(\bm{r})$ then reads
\cite{A.P.Jauho, S.Datta, B.G.Wang}
\begin{equation}
\rho = -\frac{i}{2\pi} \int d \varepsilon G^< (\varepsilon).
\end{equation}
The final density matrix is self-consistently reached from an initial guess, for which we use
the neutral atom density matrix obtained by assuming no interaction between atoms.
In the linear response region where the bias between the electrodes is small, the Fisher-Lee
relation \cite{D.S.Fisher} connects the Green's functions with the transmission coefficients.
With the density matrix and the Green's functions known,
the transmission $T(\varepsilon)$ for the quantum transport between the electrodes is then given
by \cite{S.Datta}
\begin{equation}
T(\varepsilon) = Tr[\Gamma_L G^R \Gamma_R G^A],
\end{equation}
where $\Gamma_L$, $\Gamma_R$ are line-width functions of left and right
electrodes which indicate the corresponding coupling strength between
the electrodes and central region. The linear zero-temperature conductance
is determined by the transmission at Fermi energy
\begin{equation}
G=(2e^2/h)T(E_{F}).
\label{cond}
\end{equation}

\subsection{Computational Method}

Atomistix ToolKit (ATK) \cite{Atomistix} for two-probe systems
together with a multigrid Poisson solver \cite{W.L.Briggs,T.L.Beck}
is used to carry out the numerical calculation. We at first apply
the periodic boundary condition in the $x$ and $y$ directions and
solve the corresponding electron density $\rho_0(x,y,z)$ and
electrostatic potential $V_0(x,y,z)$. $V_0(x,y,z)$ on the boundary
is denoted as the boundary condition at $V_g=0$. The boundary
condition at finite $V_g$ is obtained by shifting $V_0(x,y,z)$ with
$V_g$ on the gate but keep the same as $V_0(x,y,z)$ on the
electrodes. On each boundary region between a electrode and a gate
the potential shift decreases linearly from $V_g$ to zero. The
modified boundary condition is then applied to the system to solve
$\rho(x,y,z)$ and $V_(x,y,z)$ at arbitrary $V_g$. In the calculation
we use the SZP, single $\zeta$ valence $s$ and $p$ orbitals ($s$
orbitals) plus single $\zeta$-polarization $d$ ($p$) orbitals for Si
(H), real space SIESTA LCAO basis set \cite{Atomistix, P.Ordejon,
J.M.Soler} and the standard nonlocal norm-conserving pseudopotential
\cite{D.R.Hamann} from the database in ATK. The reason why we use
the SZP basis set instead of a simpler one is because the result
presented in this paper converges for SZP and more complete basis
sets. The local density approximation (LDA) with the Perdew-Zunger
parametrization \cite{J.P.Perdew} of the correlation energy for a
non spin-polarized homogeneous electron gas \cite{D.M.Ceperley} is
used for the exchange-correlation functional. The mesh cutoff is set
to $4348.48$ eV corresponding to a grid size of $0.092$\AA $\times
0.092$\AA $\times 0.092$\AA. The MonckHorst-Pack k-point grid is
$(1,1,100)$ along $(x,y,z)$ directions.

\section{Results and Discussions}

To express a physical property $A$ in our microscopic model on the
same footing as its counterpart in traditional macroscopic model, we
measure it from its value at $V_g=0$ and use its average over the
$y$-$z$ range inside the central unit cell ($8.15$ \AA\ $<z< 13.58$
\AA), $\langle A \rangle_{y,z}$, in the following analysis. The
total net charge $Q$ is the total charge transferred into the
supercell when $V_g$ is applied. Note that there may be a shift
between $V_g$ in our model and the observed in experiments because
i) the real boundary condition at $V_g=0$ may differ from the
periodic (bulk) boundary condition used in the calculation, ii) the
boundary potential shift when a finite $V_g$ being applied is
determined selfconsistently and can vary in different environments,
and iii) no workfunction is specified for the gate metal.

Equivalent capacitive circuits are widely used in qualitative
analysis of electrostatic properties of MISFETs in the literature.
Our system is macroscopically symmetric about the $y$-$z$ (top-down)
and $x$-$y$ (left-right) planes and three capacitors, $C_i$,
$C_{sz}$ and $C_{sx}$ may be used for this purpose. As shown in
Fig.~\ref{fig1}(b), left circuit, $C_i$ accounts qualitatively for
the average voltage drop across the insulator slab while $C_{sx}$
for the one across the Si slab in $x$ direction and $C_{sz}$ in $z$
direction. For long channel as usually exists in traditional DG
MISFETs, one-dimensional approximation is valid and only $C_{sx}$ is
important for the electrostatic analysis. For short devices,
however, the boundary at the ends of the channel also plays an
important role in determining the potential profile in the center of
the channel and the effect of $C_{sz}$ should be taken into account.
Because only the linear region is concerned in this paper, we can
also use a single capacitor $C_s$ to describe the electrostatic
properties of the Si slab with $C_s=C_{sx}
C_{sz}/(C_{sx}+C_{sz})\sim C_{sx}$, as illustrated in the right
circuit of Fig.~\ref{fig1}(b). The total net charge in the Si slab
reads
\begin{equation}
Q=V_gC=(V_g-V_s) C_i=(V_s-V_c) C_{sx}=V_c C_{sz},
\label{eq_q}
\end{equation}
with the total capacitance $C=[1/C_i+1/C_{sx}+1/C_{sz}]^{-1}$. Here $V_s$ is the surface
potential and $V_c$ the central potential of the channel as further specified in Sec.III.B.

\begin{figure}
\begin{picture}(300,180)
\put(0,0){\includegraphics{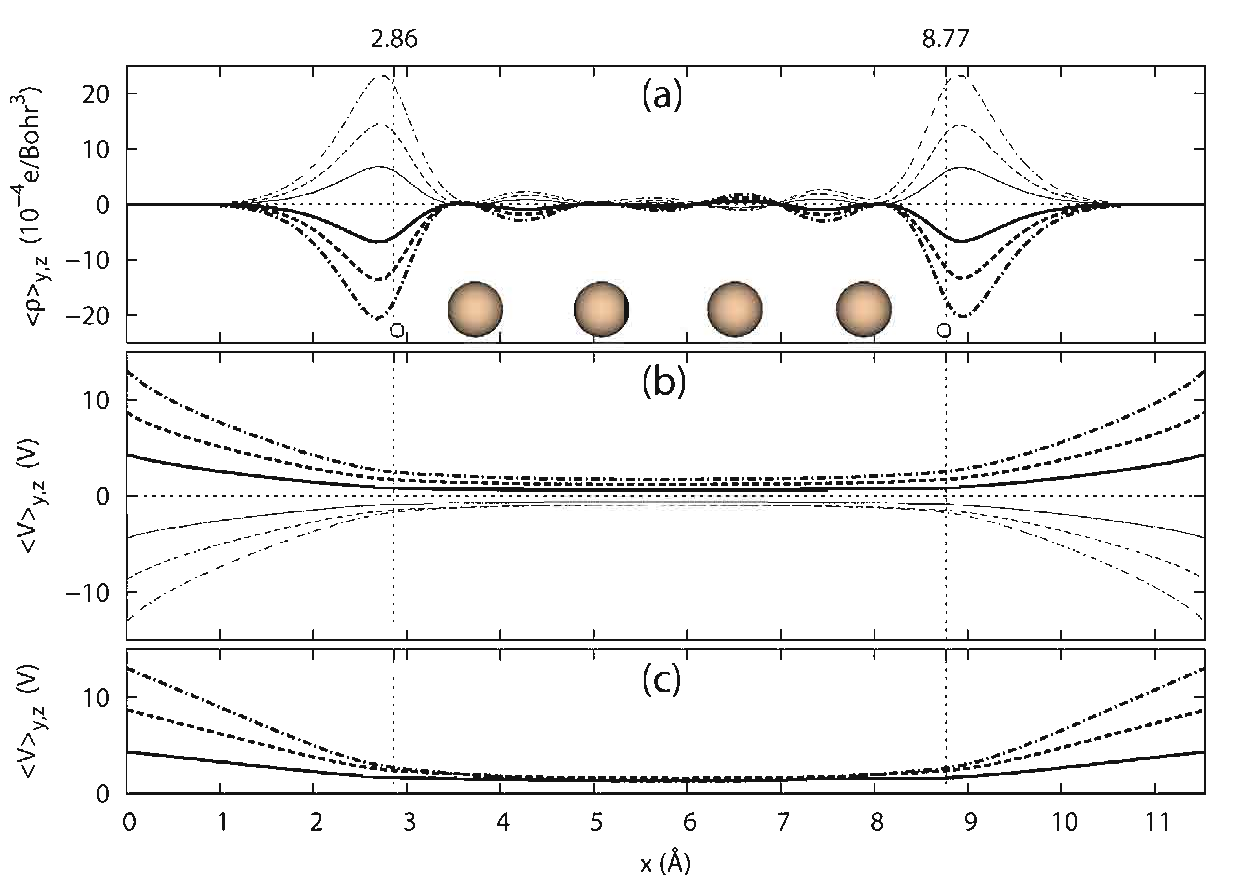}}
\end{picture}
\protect\caption{ (a) The charge density $\langle \rho\rangle
_{y,z}$ versus the $x$ coordinate for model $S_{12}$. (b) The
electrostatic potential $\langle V\rangle _{y,z}$ versus the $x$
coordinate for model $S_{12}$. (c) The same as (b) for model
$L_{12}$. Curves at positive (thick) and negative (thin) gate
voltages $V_g$ of values $0.0$, $4.35$, $8.70$, and $13.05$ V are
plotted as the dotted, solid, dashed, and dash-dotted respectively.
The vertical dashed lines at $x=2.86$ and $8.77$ \AA\ indicate the
positions of the two surfaces between the device channel and the
insulator slabs. The positions of atomic monolayers are denoted by
the big (Si) and small (H) spheres. The average $\langle \dots
\rangle _{y,z}$ is done over $y$ and $z$ in the region $0<y<5.43$
\AA\ and $8.15<z<13.58$\AA. } \label{fig2}
\end{figure}

\subsection{Charge and Potential Distributions in Real Space}

The charge density $\langle \rho\rangle_{y,z}$ distribution along
the gate ($x$) direction is shown in Fig.~\ref{fig2}(a) for model
$S_{12}$. Similar distribution is observed for other models and is
not plotted here. Attracted by the charge in the metal gates, most
of the net charges appear near the surface of the Si slab. As a
result, as illustrated in Figs.~\ref{fig2}(b) and (c), the electric
field concentrates mainly in the vacuum slab where $\langle
V\rangle_{y,z}$ varies rapidly and is weak inside the Si slab where
$\langle V\rangle_{y,z}$ becomes flat. The FET is in its off region
at small $V_g$ and $\langle V\rangle_{y,z}$ shifts with $V_g$ until
$V_g$ reaches the threshold voltage where $\langle V\rangle_{y,z}$
saturates in the center of the Si slab.

An interesting short gate effect is observed when comparing the
potential distribution for the short gate model $S_{12}$ and that
for the long gate model $L_{12}$. For the short gate model $S_{12}$,
as shown in Fig.~\ref{fig2}(b), $\langle V\rangle_{y,z}$ versus $x$
in the vacuum regions ($0<x<2.86$\AA\ or $8.77$\AA\ $<x<11.63$\AA)
is nonlinear. This is a result of the fact that the length of the
gate in model $S_{12}$ is comparable to the thickness of the vacuum
slabs in model $S_{12}$. The electric field in the vacuum slabs
deviates greatly from along the $x$ direction and the parallel-plate
capacitor model fails. For longer gate model $L_{12}$, the fringe
effect on the channel region is reduced and the $\langle
V\rangle_{y,z}$ versus $x$ curves in the vacuum slabs become
straight as shown in  Fig.~\ref{fig2}(c). The short gate effect
described above is unique in nanoscale MISFET and does not appear in
conventional MISFET since there the insulator slab's thickness is
usually much smaller than its length. We will further discuss its
effect on the FET's performance in Fig.~\ref{fig3}(c) and (d).

\begin{figure}
\begin{picture}(300,180)
\put(0,180){\includegraphics{Fig3_reduced.eps}}
\end{picture}
\protect\caption{The total net charge $Q$ in (a), (c), and (e) and
the surface potential $V_s$ in (b), (d), and (f) versus the gate
voltage $V_g$ are shown for models $S_{11}$ (solid curve), $S_{12}$
(dashed), $S_{21}$ (dotted), and $L_{12}$ (dash-dotted). Results for
models of different insulator thickness are compared in (a) and (b),
of different gate length in (c) and (d), and of different Si slab
thickness in (e) and (f). The short bars beside the curves in (b),
(d), and (f) indicate the $V_s$ values at the threshold points where
the $V_s$ versus $V_g$ curves change their slopes. } \label{fig3}
\end{figure}

\subsection{Total Transferred Charge and Surface Potential}

We use the surface just enclosing the H and Si atoms as indicated by
the horizontal dashed lines in Fig.~\ref{fig1}(a) or the vertical
dashed lines at $2.86$ \AA\ and $8.77$ \AA\ for model $S_{12}$ in
Fig.~\ref{fig2}(a) as the surfaces separating the channel and the
insulator slab. The averaged value $\langle V\rangle_{y,z}$ on these
surfaces is used as the surface potential $V_s$. The central
potential $V_c$ is approximated as the $\langle V\rangle_{y,z}$
value in the middle of the two gates, i.e. at $x=5.82$ \AA\ in
Fig.~\ref{fig2}(a) for model $S_{12}$.

With $V_s$ and $V_c$ known at each $V_g$, we can estimate the
capacitance from Eq.~(\ref{eq_q}). For traditional MISFETs, usually
we have $C_{sx} < C_i$ in the off region and $C_{sx} > C_i$ in the
on region. However, the Si slab here is extremely thin and $C_{sx} >
C_i$ in both regions. In addition, the span of the device in the $z$
direction is larger than that in the $x$ direction. This fact leads
to $C_{sx} > C_{sz}$ in the off-region and $C_s$ may increase with
the Si slab thickness because $C_{sz}$ is proportional to the
thickness while $C_{sx}$ is proportional to the inverse of it. In
the on region, when a large amount of free charges are created on
the surface, $V_c$ is almost pinned due to the screening effect of
these free charges.

The total net charge $Q$ versus the gate voltage $V_g$ for model
$S_{12}$ is shown by the dashed curve in Fig.~\ref{fig3}(a). At
$V_g=0$, the Fermi energy $E_F$ is located in the energy gap of the
Si slab. As $V_g$ varies, the relative position of the Fermi level
in the energy band shifts accordingly. $C_s$ remains almost constant
when the Fermi level is in the energy gap but increases quickly as
the Fermi level enters the conduction or the valence band. This
sudden increase of $C_s$ happens at the positive and negative
threshold points, i.e., near $V_g=13$ V and $-7$ V, respectively.
Near these two points, due to the shift of the $C_s/C_i$ ratio, the
$V_s$ versus $V_g$ curve also changes its slope as shown by the
dashed curve in Fig.~\ref{fig3}(b).

To show the effect of the vacuum-slab thickness on the electrostatic
characteristics, we use model $S_{11}$ with a thinner vacuum slab
and redo the calculation. $\langle \rho\rangle _{y,z}$ and $\langle
V\rangle _{y,z}$ have similar profiles between the gates as that for
model $S_{12}$ but with a larger amplitude at the same gate voltage.
In the equivalent capacitive circuit, $C_i$ increases with the
shrinking of the vacuum slab, which results in the shift of $Q$ and
$V_s$ as shown by the solid curve in Fig.~\ref{fig3}(a) and (b). The
$Q$ versus $V_g$ curve of thicker insulator slab model $S_{12}$ has
a wider subthreshold region in the $V_g$ axis. One interesting
observation is that the threshold surface potentials, the surface
potentials at the threshold points, for model $S_{11}$ and $S_{12}$
are very close as indicated by the short bars beside the curves in
Fig.~\ref{fig3}(b). This is because the surface potential reflects
directly the Fermi level in the energy band of the Si slab. For a
system of thicker insulator or smaller $C_i$ it takes a bigger gate
voltage difference to shift the Fermi energy from the valence band
to the conduction band.

In Fig.~\ref{fig2} we have shown that the electric field in the
insulator slab can deviate from normal to the slab due to the fringe
effect in short gate cases. This short gate effect in nano FET
reduces the per area capacitance of the insulator slab and results
in a larger threshold gate voltage than estimated from parallel
capacitance approximation. In Fig.~\ref{fig3}(c) and (d) we plot the
$Q$ versus $V_g$ curves and the $V_s$ versus $V_g$ curves
respectively for model $S_{12}$ (dashed) and longer gate model
$L_{12}$ (dash-dotted). The smaller total capacitance $C$ in model
$S_{12}$, due to its half gate length plus the short gate effect in
it, results in a less than half net charge $Q$ in the system at any
$V_g$ and a much gentler slope of its $Q$ versus $V_g$ curve in
Fig.~\ref{fig3}(c). In Fig.~\ref{fig3}(d), we also observe a gentler
slope of the $V_s$ versus $V_g$ curve for model $S_{12}$ in the
subthreshold region. Combining the relation $V_s/V_g=C_i/(C_i+C_s)$
as expressed in Eq.~(\ref{eq_q}), this means a smaller ratio of
$C_i/C_s$ and confirms the conclusion drawn from Fig.~\ref{fig2}
that the short gate effect mainly reduces the insulator capacitance.

Up to now, we have seen that both making the insulator slab thinner
and suppressing the short gate effect can enhance the insulator
capacitance and reduce the threshold gate voltage as illustrated in
Fig.~\ref{fig3}(b) and (d). On the other hand, a close comparison of
the curves shows that the two changes to the system have opposite
effects on the threshold surface potential. The threshold surface
potential has a tendency to increase in the former case (the solid
bar on the positive $V_g$ side in Fig.~\ref{fig3}(b) is slightly
higher than the dashed one) while the threshold surface potential
decreases in the latter case. A detailed study shows that the
variation of the threshold surface potential is related to the
potential distribution inside the Si slab. Given the same surface
potential in the subthreshold region, the potential distribution
along $x$ changes little when varying the insulator slab thickness
but a potential drop appears and broadens in the center of the
$\langle V\rangle_{y,z}$ versus $x$ curves when shortening the gate.

In Fig.~\ref{fig3}(e) and (f), we present the dependence of
electrostatic characteristics on the Si slab thickness exemplified
by model $S_{11}$ and $S_{21}$. Similar to the case of thicker
insulator slab as shown in Fig.~\ref{fig3}(a) and (b), the $Q$ and
$V_s$ versus $V_g$ curves for a system of thicker Si slab also have
a wider subthreshold region in the $V_g$ axis. However, this
apparent common feature in the two cases has different origins and
it is interesting to compare the details of their electrostatic
characteristics.

At first, as pointed out earlier in this section, because the Si
slab inside the supercell has a dimension in the $z$ direction
bigger than that in the $x$ direction, $|C_{sx}| > |C_{sz}|$ and
$C_s$ increases with the Si slab thickness in the off-region. In
this case, the $Q$ versus $V_g$ curve of thicker Si slab model
$S_{21}$ (dotted) in Fig.~\ref{fig3}(e) has a sharper slope in the
off-region due to an increased total capacitance $C$, in contrast to
the gentler slope of the curve for thicker insulator slab model
$S_{12}$ (dashed) in Fig.~\ref{fig3}(a). In addition, the bigger
$C_s$ in the thicker Si slab case, in stead of the smaller $C_i$ as
in the thicker insulator slab case, results in the gentler slope of
the dashed $V_s$ versus $V_g$ curve in Fig.~\ref{fig3}(f).

Secondly, it is well known \cite{Y.Omura,P.V.Sushko,S.Datta.1} that
the energy gap of semiconductor materials in nanostructures is
enlarged from their bulk value as a result of the quantum
confinement effect. The change of the gap depends on the confining
potential profile characterized by parameters such as the confining
dimension and the potential height. This effect is automatically
taken into account by our NEGF-DFT model. As will be illustrated
later in Sec.III.C, the thicker Si slab in model $S_{21}$ has a
narrower and shifted energy gap compared to that in model $S_{11}$.
As a result, the subthreshold region in the $V_s$ axis, i.e. the
difference between the surface potentials at the positive and
negative threshold points, of model $S_{21}$ ($3.79$ V for the
dotted curve in Fig.~\ref{fig3}(f)) is narrower than that of model
$S_{11}$ ($3.99$ V for the solid curve). Furthermore, we also
observe a shift of the subthreshold region in the $V_s$ and $V_g$
axes when the Si slab thickness varies.

As discussed above and illustrated in Fig.~\ref{fig3}(e) and (f),
the thicker Si slab in model $S_{21}$, compared to that in model
$S_{11}$, narrows the subthreshold $V_s$ region via the quantum
confinement effect on the one hand and reduces $V_s/V_g$ ratio in
the off-region due to the specific length-thickness ratio of the Si
slabs in the models on the other hand.  The observed wider
subthreshold $V_g$ region is a result of the competition between
these two independent but opposite factors.

In this section, we analyze the electrostatic characteristics of our
model nano FET in a way widely employed in the analysis of
conventional MISFET. \cite{N.Arora} Here we want to emphasize that
our system (of size 2 nm) is much smaller than any of the
conventional MISFETs even the smallest predicted by ITRS in its
latest report. \cite{roadmap} As a result, our model device shows
characteristics between molecular electronics and conventional
electronics. For example, we observe a large slope of the $Q$ versus
$V_g$ curve in the off region in all models which is similar to the
result of the Si$_4$ cluster \cite{Z.X.Dai} but different from that
in conventional MISFET where a flat $Q$-$V_g$ dependence is usually
found in the off region. In addition, the threshold gate voltage in
our systems ($> 5$ V) is much higher than the value usually observed
in conventional MISFET ($< 2$ V). This is because the Si slab in our
systems is very thin and a vacuum instead of SiO$_2$ slab is used as
the insulator slab. The $C_s/C_i$ ratio then becomes much larger in
the off region.

\begin{figure}
\begin{center}
\begin{picture}(300,180)
\put(0,180){\includegraphics{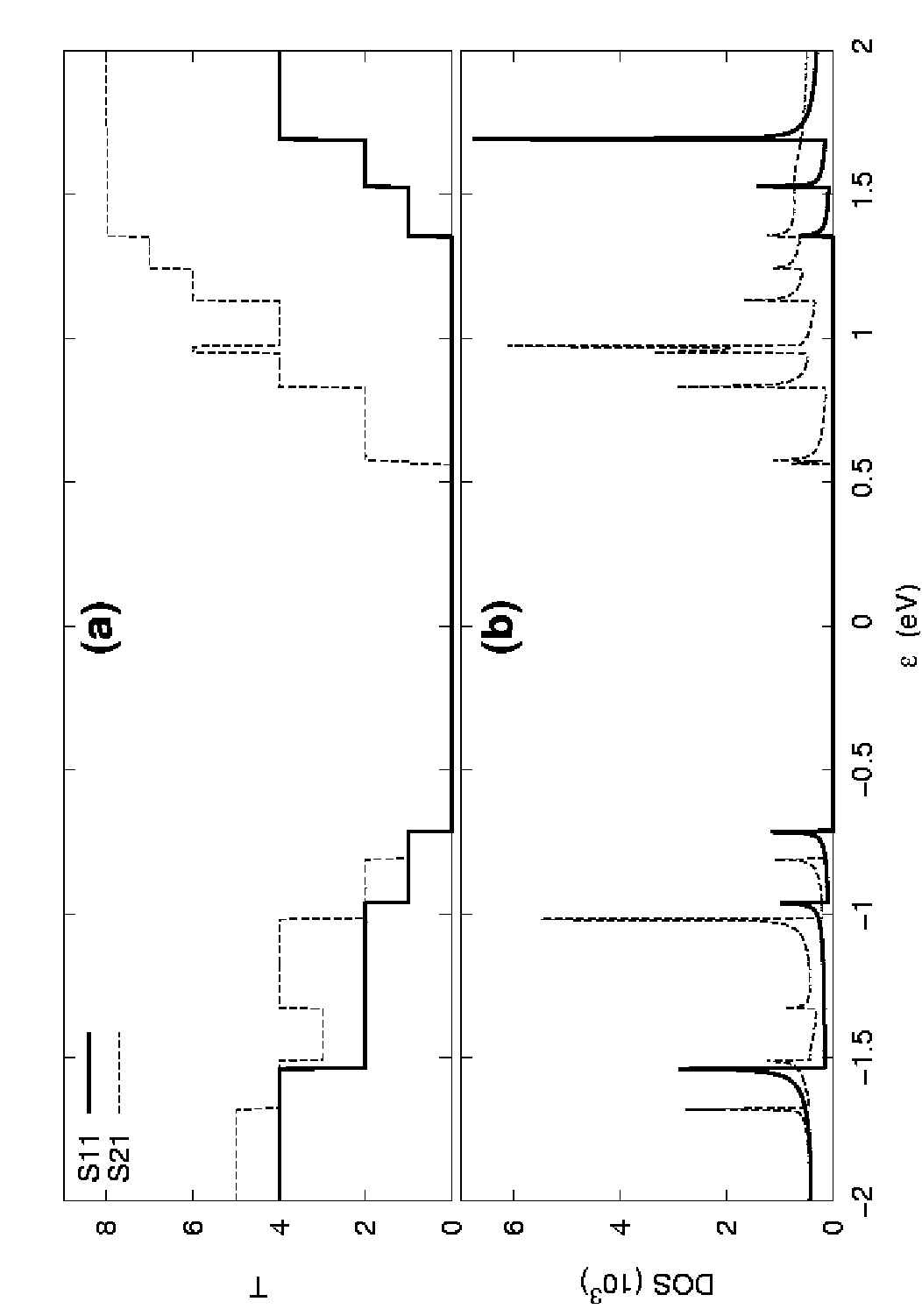}}
\end{picture}
\protect\caption{ (a) The transmission $T$ and (b) the density of
states (DOS) for electrons of momentum $(0,0,k_z)$ are plotted as
functions of the electron energy $\varepsilon$ in models $S_{11}$
(solid curve) and $S_{21}$ (dashed).} \label{fig4}
\end{center}
\end{figure}

\subsection{Transmission and Density of States}

To explore further the origin of the electrostatic properties of
nano FETs and the gate effect on transport, we calculate the
transmission, which gives the conductance in the linear region via
Eq.~(\ref{cond}), and the density of states (DOS). In
Fig.~\ref{fig4}(a), the transmission of an electron propagating
along the $z$ direction is calculated at $V_g=0$ for quantum
transport between the source and drain electrodes in model $S_{11}$
(solid) and $S_{21}$ (dashed). Because the electron sees a perfect
crystal in the $z$ direction, its transmission is $100\%$ for each
channel and the transmission versus energy curve tells the number of
propagating channels for an electron of the energy. The
one-dimensional DOS in $k_z$ direction for model $S_{11}$ (solid)
and $S_{21}$ (dashed) is plotted in Fig.~\ref{fig4}(b). The band gap
of model $S_{11}$ is wider than that of model $S_{21}$ due to the
quantum-confinement effects in nano semiconductor structure
\cite{S.Datta.1}. In addition, most probably due to the change of
the surface effect or the ratio between the numbers of the H and Si
atoms in the system, a shift of the energy gap is observed when
varying the Si slab thickness.

\begin{figure}
\begin{picture}(300,150)
\put(0,180){\includegraphics{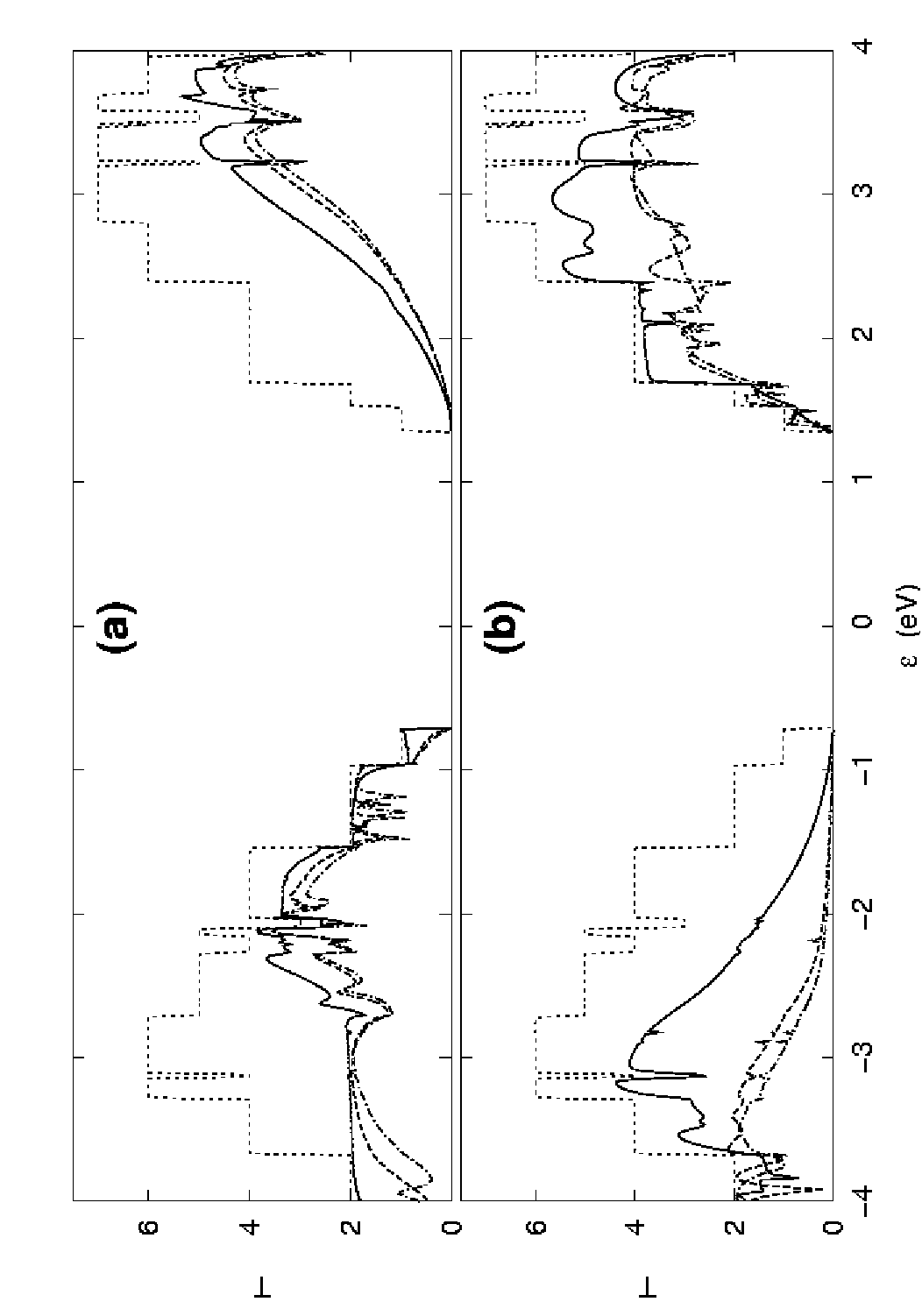}}
\end{picture}
\protect\caption{ The transmission $T$ versus energy $\varepsilon$
at $|V_g|=0.0$ (dotted), $4.35$ (solid), $8.70$ (dashed), and
$13.05$ V (dot-dashed) is illustrated for negative $V_g$ (a) and
positive $V_g$ (b). Model $S_{11}$ is used in the calculation.}
\label{fig5}
\end{figure}

The transmission versus electron energy at different gate voltages
is plotted in Fig.~\ref{fig5} for model $S_{11}$. When a negative
gate voltage is applied, the energy band in the gate region shifts
to higher energy. This shift forms an energy barrier for electrons
and a well for holes. As a result, we observe a significant decrease
of transmission for electrons near the bottom of the conduction band
as shown in Fig.~\ref{fig5}(a). On the contrary, if a positive gate
voltage is applied, an energy barrier is formed for the holes in the
middle of the device and the propagation of holes near the top of
the valence band becomes unfavored. The transmission profile changes
quickly with the gate voltage in the subthreshold region and then
becomes less sensitive when $V_c$ and $V_s$ saturate at $V_g$ higher
than the threshold gate voltage.

\section{summary}

In summary, we have proposed an atomistic multi-terminal model for
electrostatic simulation of nano MISFETs in the framework of the
NEGF-DFT approach and studied the behaviour of devices with Si slabs
of one and two unit cells in thickness, with variable insulator
(vacuum in this model) thickness, and with different gate length.
The subthreshold region in terms of the surface potential reflects
the band gap of the Si slabs and the surface potential at the
threshold points remains little changed in devices of different
insulator thickness. For a device with thicker Si slab, the
subthreshold region in surface potential becomes narrower due to the
quantum confinement effect while the subthreshold region in gate
voltage becomes wider as a result of the competition between the
confinement effect and the geometry effect on capacitance. In nano
FETs where the gate length is comparable to the insulator thickness,
the short gate effect results in a much wider subthreshold region
than that observed in conventional MISFETs. This short gate effect
can be suppressed by using longer gates to reduce the fringe effect
on the electric field in the insulator slab.

In addition, we have calculated the band structure of Si slabs with
different thickness. The density of states and the transmission
probability of electrons in our nano FET system are then estimated
to demonstrate the quantum confinement effect and the gating effect
on transport in the linear region. The energy gap widens and shifts
to higher energy when the Si slab varies from one to two unit-cells
in thickness. The application of a gate voltage introduces energy
barriers or wells to the carriers in the transport channel and
modifies the performance the nano FETs. A negative gate voltage
favors the hole transport between the electrodes while a positive
one favors the electron transport.

Limited to the computation capability, our model system is still not fully realistic and is
smaller than even the latest feasible MISFETs \cite{roadmap}. Nevertheless, our model may
provide qualitative information about realistic MISFETs and serve as a prototype framework for
nano MISFET simulations.

\begin{acknowledgments}
This work has been supported in part by the Economic Development Board
(EDB), Singapore, and in part under the joint Research Collaboration
Agreement between Nanyang Technological University (NTU) and Atomistix
Asia Pacific Pte Ltd (AAP).  Support from the Nanocluster and
Microelectronics Center of NTU is also acknowledged.
\end{acknowledgments}


\begin{thebibliography}{99}
\bibitem{note} Approaches of the DFT and NEGF combination used in the literature can be
different depending on the assumptions, the approximations, and the basis set used in specific
cases. In the paper, we specify the approach we used in this work and refer the readers to the
related references for detailed analysis and justification of the approach since it has been
systematically described elsewhere.
\end{thebibliography}
\end{document}